\documentclass[letter]{aa}

\usepackage[varg]{txfonts}
\usepackage{graphicx}
\usepackage[breaklinks=true,colorlinks,linkcolor=blue,citecolor=blue,urlcolor=blue]{hyperref}

\newcommand{\norm}[1]{\left\lVert#1\right\rVert}
\newcommand{\ve}[1]{{\vec{#1}}}

\providecommand{\sorthelp}[1]{} % For Planck_bib.bib

\graphicspath{{figures/}}

\begin{document}

\title{A new approach for the statistical denoising of \textit{Planck} interstellar dust polarization data\thanks{The public Python package \texttt{PyWPH} is available at \url{https://github.com/bregaldo/pywph}.}}

\subtitle{}

\author{Bruno Regaldo-Saint Blancard\inst{\ref{lpens},\ref{obs}}
\and Erwan Allys\inst{\ref{lpens}}
\and François Boulanger\inst{\ref{lpens}}
\and François Levrier\inst{\ref{lpens}}
\and Niall Jeffrey\inst{\ref{lpens}, \ref{ucl}}}

\institute{{Laboratoire de Physique de l'École Normale Supérieure, ENS, Université PSL, CNRS, Sorbonne Université, Université de Paris, 75005 Paris, France\\ \email{bruno.regaldo@phys.ens.fr}\label{lpens}} \and {Observatoire de Paris, PSL University, Sorbonne Université, LERMA, 75014 Paris, France\label{obs}} \and {Department of Physics \& Astronomy, University College London, Gower Street, London, WC1E 6BT, UK\label{ucl}}}

\date{Received: February 5, 2021 /
Accepted: May 13, 2021}

\abstract
% Context
% Aims
% Methods
% Results
% Conclusions
{Dust emission is the main foreground for cosmic microwave background (CMB) polarization. Its statistical characterization must be derived from the analysis of observational data because the precision required for a reliable component separation is far greater than what is currently achievable with physical models of the turbulent magnetized interstellar medium. This letter takes a significant step toward this goal by proposing a method that retrieves non-Gaussian statistical characteristics of dust emission from noisy {\it Planck} polarization observations at $353\,$GHz. We devised a statistical denoising method based on wavelet phase harmonics (WPH) statistics, which characterize the coherent structures in non-Gaussian random fields and define a generative model of the data. The method was validated on mock data combining a dust map from a magnetohydrodynamic simulation and {\it Planck} noise maps. The denoised map reproduces the true power spectrum down to scales where the noise power is an order of magnitude larger than that of the signal. It remains highly correlated to the true emission and retrieves some of its non-Gaussian properties. Applied to {\it Planck} data, the method provides a new approach to building a generative model of dust polarization that will characterize the full complexity of the dust emission. We also release \texttt{PyWPH}, a public Python package, to perform GPU-accelerated WPH analyses on images.
}


\maketitle

%%%%%%%%%%%%%%%%%%%%%%%%%%%%%%%%%%%%%%%%%%%%%%%%%%%%
%%%%%%%%%%%%%%%%%%%%%%%%%%%%%%%%%%%%%%%%%%%%%%%%%%%%
%% SECTION 1: INTRODUCTION
%%%%%%%%%%%%%%%%%%%%%%%%%%%%%%%%%%%%%%%%%%%%%%%%%%%%
%%%%%%%%%%%%%%%%%%%%%%%%%%%%%%%%%%%%%%%%%%%%%%%%%%%%

\section{Introduction}

The quest of primordial B-modes in the polarization of the cosmic microwave background (CMB) faces a major challenge, namely the accurate characterization of one of its main foregrounds: the polarized emission of interstellar dust from our Galaxy \citep{Ade2015a}. To rightfully claim any detection of the cosmological signal, one has to fully take into account the complexity of the Galactic contribution to the sky emission~\citep{Hensley2018, Clark2019,Pelgrims2020,planck2016-l11A}. The statistical characterization of Galactic polarized foregrounds is also important to extract the CMB lensing potential from sky observations \citep{Beck_2020}.

The Galactic contribution is sourced by the thermal emission of nonspherical dust grains in the diffuse interstellar medium (ISM). The mixture of dust and gas is described as a turbulent magnetized fluid, and the grains tend to align statistically with their long axes perpendicular to the local magnetic field, thus polarizing the emission \citep{planck2016-l11B}. The physical and chemical processes regulating and structuring the diffuse ISM are nonlinearly coupled~\citep{Draine2011}, leading to the emergence of strong scale couplings, which may be evidenced by the highly non-Gaussian statistics observed for many tracers~\citep{Burkhart2009}.

Capturing non-Gaussianity is therefore essential to characterize the dust signal. To tackle this issue, several authors have introduced machine learning algorithms that need to be trained~\citep{10.1093/mnras/staa3344, Krachmalnicoff2020, Petroff2020, Thorne2021}. Another approach, which does not involve any learning steps, uses statistics defined from nonlinear transforms, namely the wavelet scattering transform and the wavelet phase harmonics (WPH). These describe the couplings between scales in non-Gaussian fields, and they have been applied to ISM data \citep{Allys2019, Regaldo-SaintBlancard2020, Saydjari2020} as well to study the large-scale structure of the Universe \citep{Allys2020, Cheng2020}. They have been successfully used to characterize simulated and observational data with a high signal-to-noise ratio (S/N), and to build generative models for these data. However, these approaches face difficulties with noisy data, which is the case of {\it Planck} polarization observations.
In this letter, we introduce a statistical denoising method using the WPH statistics to retrieve the statistical properties of the noise-free dust emission.

In Sect.~\ref{sec:method} we present our statistical denoising method and show how well it performs on noisy simulated data of the polarized emission from dust. In Sect.~\ref{sec:planck}, we apply it to the Chamaeleon-Musca 353~GHz polarized emission, as observed by the {\it Planck} satellite, and discuss our results. In Sect.~\ref{sec:conclusion} we conclude and suggest a few perspectives for follow-up studies. This paper also includes two appendices. Appendix~\ref{app:wph} introduces the WPH statistics and the denoising procedure in detail. Appendix~\ref{app:comparison} presents a first comparison with other methods. We also provide a public Python package to perform GPU-accelerated WPH analyses on images called \texttt{PyWPH}\footnote{\url{https://github.com/bregaldo/pywph}}.

%%%%%%%%%%%%%%%%%%%%%%%%%%%%%%%%%%%%%%%%%%%%%%%%%%%%
%%%%%%%%%%%%%%%%%%%%%%%%%%%%%%%%%%%%%%%%%%%%%%%%%%%%
%% SECTION 2: METHOD AND FIGURES OF MERIT
%%%%%%%%%%%%%%%%%%%%%%%%%%%%%%%%%%%%%%%%%%%%%%%%%%%%
%%%%%%%%%%%%%%%%%%%%%%%%%%%%%%%%%%%%%%%%%%%%%%%%%%%%

\section{Method and validation}
\label{sec:method}

\begin{figure*}
    \centering
    \includegraphics[width=\textwidth]{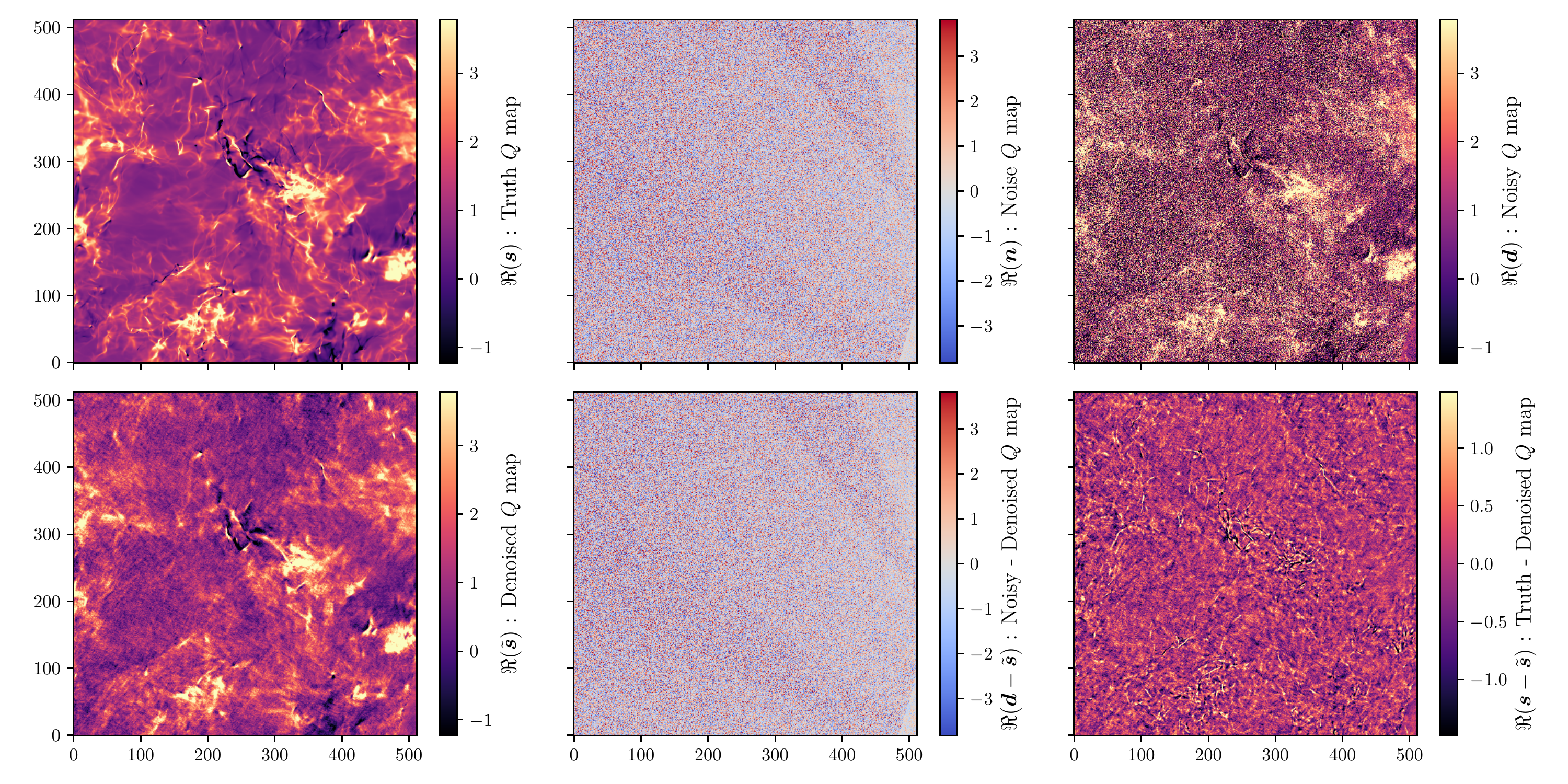}
    \caption{{\it Top row}: Simulated $Q$ maps corresponding to the truth $\ve{s}$ (left), the noise $\ve{n}$ (middle), and the resulting noisy map $\ve{d}$ (right). {\it Bottom row}: Denoised simulated $Q$ map $\Re(\ve{\tilde{s}})$ (left) next to the difference noisy-denoised $\Re(\ve{d} - \ve{\tilde{s}})$ (middle) and denoising error $\Re(\ve{s}-\ve{\tilde{s}})$ (right). Units are arbitrary, but kept consistent between the maps.}
    \label{fig:simulated_maps}
\end{figure*}

\subsection{Description of the method}
\label{sec:desc_method}

We observe a noisy map $\ve{d}$ that is modeled as follows:
\begin{equation} \label{eq:data_model}
    \ve{d} = \ve{s} + \ve{n},
\end{equation}
with $\ve{s}$ being the target truth map and $\ve{n}$ being an additive noise signal. The noise $\ve{n}$ is an unknown realization of a random field $\ve{N}$. We assume that we know $\ve{N}$, meaning that we are able to generate as many independent realizations of it as needed. We call $\{\ve{n}_1, \ldots , \ve{n}_M\}$ $M$ of these.

We introduce a method to retrieve the statistical properties of $\ve{s}$ while denoising $\ve{d}$. This statistical denoising consists in iteratively deforming $\ve{d}$ to build a denoised map $\tilde{\ve{s}}$ such that the $\{\tilde{\ve{s}} + \ve{n_i}\}_i$ maps and the $\ve{d}$ map become "close enough" in a given statistical space. It employs the possibility introduced in \citet{Zhang2019} to build a generative model from WPH statistics, that is, the ability to generate new random realizations based on statistical constraints. In the following, we call $\phi$ the operator that computes a set of summary statistics from a given map. The algorithm consists in minimizing the following loss function:
\begin{equation}
\label{eq:loss}
    {\cal L}_M(\ve{u}) = \frac1{M}\sum_{i = 1}^M\norm{\phi\left(\ve{u} + \ve{n}_i\right) - \phi\left(\ve{d}\right)}^2,
\end{equation}
where $\norm{~\cdot~}$ denotes the Euclidean norm. We choose $\ve{u}_0 = \ve{d}$ to initialize the optimizer. The denoised map $\tilde{\ve{s}}$ corresponds to an approximate minimum obtained by performing this optimization in pixel space, using an LBFGS optimizer \citep{Byrd1995}. We note that a limitation of this algorithm relies on the (ideal) assumption that we know $\ve{N}$. In practice, any non-modeled statistical property of the noise will be considered to be part of the signal $\ve{s}$.

The choice of the operator $\phi$ is obviously paramount to the quality of the method and must be tailored to the properties of $\ve{s}$ and $\ve{n}$. In the context of ISM polarization data, we expect $\ve{s}$ to be relatively regular and to exhibit non-Gaussian signatures due to the interactions between scales (e.g., filamentary structures), while the noise is expected to be highly irregular and close to a (possibly spatially-varying) Gaussian white noise.

In this work, the $\phi$ operator computes WPH statistics. These statistics are designed to characterize the coherent structures that appear in non-Gaussian random fields, by quantifying the phase alignment between different scales~\citep{Zhang2019, Mallat2020}. Their ability to define a generative model has been studied in particular by~\citet{Allys2020}, who demonstrate quantitatively that they include most of the information captured by various other statistics, such as the power spectrum, the bispectrum, or the Minkowski functionals. These statistics were applied to real-valued data. Here we extend them to complex-valued polarization maps $Q+iU$. Statistics derived from this complex variable efficiently characterize the full complexity of the polarization signal~\citep{Regaldo-SaintBlancard2020}. We refer readers to Appendix~\ref{app:wph} for a presentation of the WPH statistics and a detailed description of the denoising procedure.

\begin{figure}
    \centering
    \includegraphics[width=0.95\hsize]{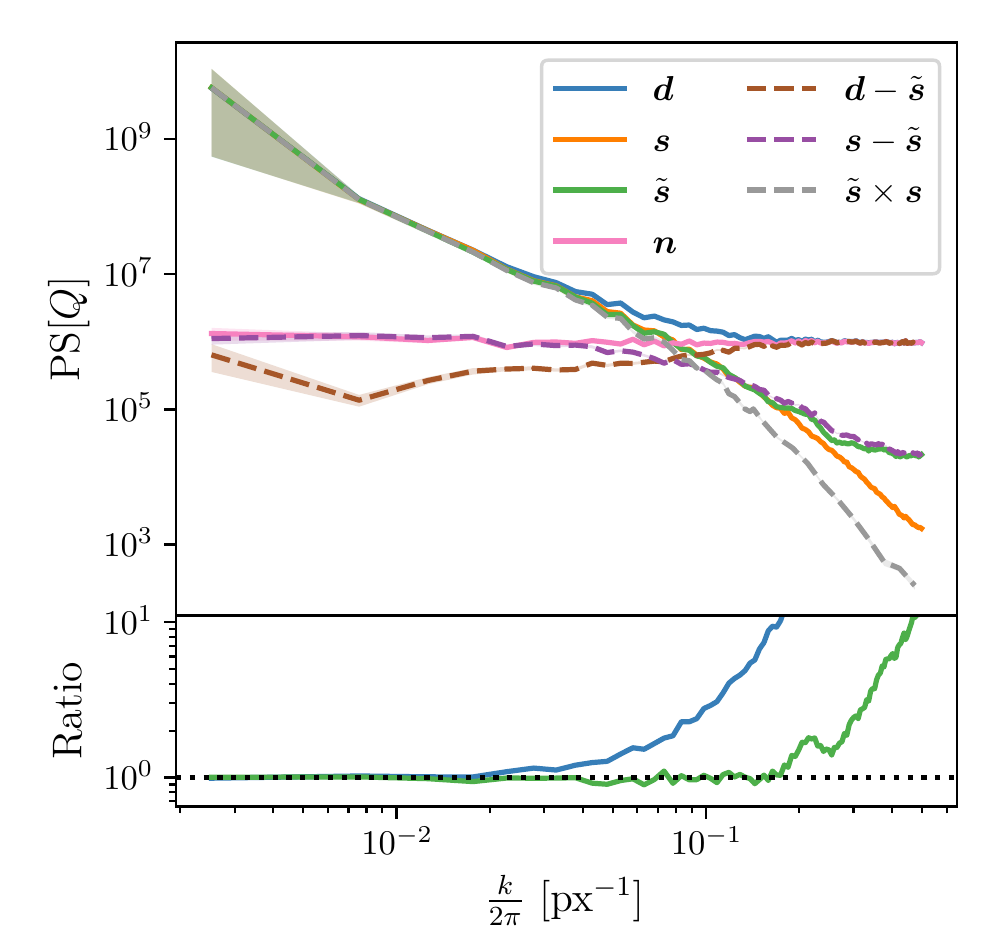}
    \caption{$Q$ maps power spectra for $\ve{d}$, $\ve{s}$, $\ve{\tilde{s}}$, $\ve{n}$ (all in solid lines), $\ve{d}-\ve{\tilde{s}}$, and $\ve{s}-\ve{\tilde{s}}$, and cross-spectrum between the $Q$ maps of $\tilde{\ve{s}}$ and $\ve{s}$ (all in dashed lines). In the bottom panel, we also show the ratio of the power spectra of $\ve{d}$ and $\ve{\tilde{s}}$ with that of $\ve{s}$.}
    \label{fig:simulated_ps}
\end{figure}

\begin{figure*}
    \centering
    \includegraphics[width=\textwidth]{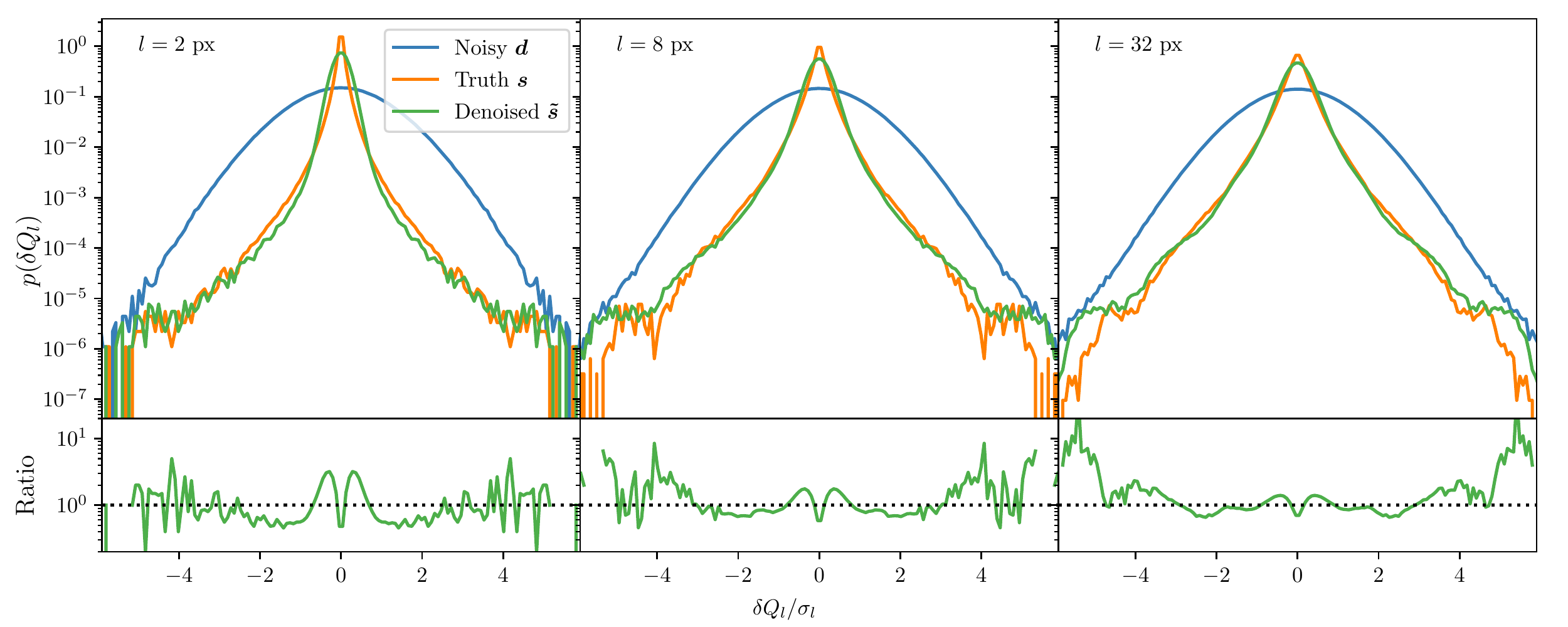}
    \caption{PDFs of the increments of $Q$ for $\ve{d}$ (noisy), $\ve{s}$ (truth), and $\ve{\tilde{s}}$ (denoised), computed for three logarithmically spaced lags going from 2 to 32 pixels. For each lag, the increments were normalized by the standard deviation $\sigma_l$ of the Gaussian that fits the core of the PDF of the noisy map. In the bottom panels, we also show the ratio of the PDFs of $\ve{\tilde{s}}$ with those of $\ve{s}$ for each lag.}
    \label{fig:simulated_increments}
\end{figure*}

\subsection{Validation on a simulation}
\label{sec:fig_merit}

In this section, we assess the performance of our denoising algorithm by applying it to mock data $\ve{d} = \ve{s} + \ve{n}$, emulating a noisy $Q+iU$ {\it Planck} polarization signal. We show figures of merit based on power spectra and probability distribution functions (PDFs) of the increments of the maps. These are only shown for $Q$, but they are similar for $U$. We postpone figures of merit based on other kinds of statistics, including the WPH statistics, to a future study.

Here $\ve{s}$ is a simulated $Q+iU$ map, representing a typical linear polarization signal produced by the thermal emission of dust in the diffuse ISM. It was built from the same magnetohydrodynamic simulation as the one used in \citet{Regaldo-SaintBlancard2020} and following the same procedure. We refer to this paper for more details on the construction of this simulated map. We simulated a {\it Planck} observation of dust polarization using {\it Planck} instrumental noise maps introduced in Sect.~\ref{sec:planck}~\citep{planck2016-l03}. We have a total of 300 realizations of this noise at our disposal. We picked one of these, called $\ve{n}$, to build $\ve{d}$, and we used the remaining $M=299$ noise maps, labeled $\{\ve{n}_1, \dots, \ve{n}_M\}$, for the denoising algorithm.

We define the S/N of $\ve{d}$ as the ratio of the standard deviations of $\Re(\ve{s})$ and $\Re(\ve{n})$ (in the following, the real part $\Re$ of a complex $Q+iU$ map refers to its $Q$ map). We adjusted this S/N by scaling $\ve{s}$, so that the impact of the noise "resembles" that on the {\it Planck} map presented in Sect.~\ref{sec:planck}. This was not straightforward since the power spectrum of the simulated map $\ve{s}$ has a different slope than that of the estimate of the power spectrum of the dust emission from the {\it Planck} map (see Figs.~\ref{fig:simulated_ps} and \ref{fig:observed_ps}). We decided to adjust the S/N so that the scale $k_i$, at which the power spectra of $\ve{n}$ and $\ve{s}$ intersect, coincides with the one from the {\it Planck} map. Figs.~\ref{fig:simulated_ps} and \ref{fig:observed_ps} show that ${k_i/(2\pi)\approx 0.8~\text{px}^{-1}}$. This procedure leads to $\text{S/N} \approx 0.4$ for the simulated map, which is more than two times lower than the S/N of the observational map discussed in Sect.~\ref{sec:planck}.

\begin{figure*}
    \centering
    \includegraphics[width=0.95\textwidth]{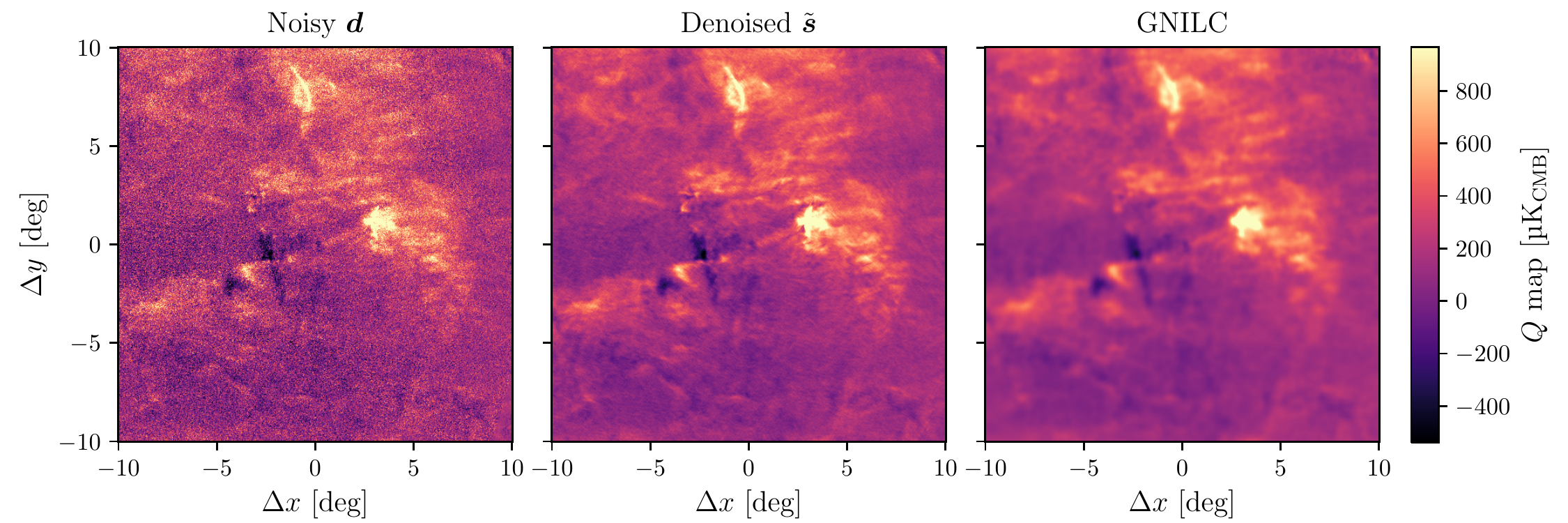}
    \caption{Noisy (left) and denoised (middle) $Q$ maps of the Chamaeleon-Musca region as observed by the {\it Planck} satellite at 353~GHz. The noisy maps were denoised as described in Sect.~\ref{sec:desc_method}. We also show the corresponding GNILC map (right) for reference.}
    \label{fig:observed_maps}
\end{figure*}

We show in Fig.~\ref{fig:simulated_maps} (top) the simulated $Q$ maps corresponding to the truth $\ve{s}$ (left), the noise $\ve{n}$ (middle), and the resulting noisy map $\ve{d}$ (right). The map $\Re(\ve{s})$ exhibits coherent structures such as filaments and large smooth regions that are characteristic of its non-Gaussian statistical properties. On the other hand, $\Re(\ve{n})$ seems close to an inhomogeneous white Gaussian noise. The spatial inhomogeneity appears as variations of the local standard deviation due to the {\it Planck} satellite scanning strategy. Finally, in $\Re(\ve{d})$, coherent structures at intermediate and small scales are hard or impossible to identify.

We applied our denoising method to the simulated noisy map $\ve{d}$ and show, in Fig.~\ref{fig:simulated_maps} (bottom), the resulting denoised $Q$ map $\Re(\ve{\tilde{s}})$ (left) next to the difference between the noisy and denoised maps $\Re(\ve{d} - \ve{\tilde{s}})$ (middle) and the denoising error $\Re(\ve{s}-\ve{\tilde{s}})$ (right). The map $\Re(\ve{\tilde{s}})$ shows that the noise level has been drastically reduced and that we were able to recover the filamentary structure down to a minimum scale. We can identify the smooth regions of $\Re(\ve{s})$ even if there still remains a visible noise. The similarities between $\Re(\ve{d} - \ve{\tilde{s}})$ and $\Re(\ve{n})$ are striking. The local variations of the standard deviation of the noise are clearly recovered, demonstrating that the inhomogeneity of the noise is not an issue for our method. Finally, $\Re(\ve{s} - \ve{\tilde{s}})$ exhibits some remaining structures that match the thinnest filaments appearing in $\Re(\ve{s})$, on top of a more diffuse background. This indicates that down to a minimum scale, below which the algorithm struggles to recover features, most of the structures are efficiently reconstructed.

Figure~\ref{fig:simulated_ps} compares the power spectra of the six maps shown in Fig.~\ref{fig:simulated_maps} plus the cross-spectrum between the denoised and truth maps. These power spectra were estimated computing the mean square moduli of the Fourier components of the maps binned on linearly spaced isotropic wavenumber bins, and the uncertainties correspond to the standard deviation of the mean. The cross-spectrum was computed similarly except for the bins above $k/(2\pi) = 0.14~\text{px}^{-1}$ that were logarithmically spaced in order to lower the statistical variance\footnote{We recall that this cross-spectrum, called $\text{CS}[\Re(\ve{s}), \Re(\tilde{\ve{s}})]$, is related to the power spectra by the following relation: ${\text{PS}[\Re(\ve{s}-\tilde{\ve{s}})] = \text{PS}[\Re(\ve{s})] + \text{PS}[\Re(\tilde{\ve{s}})] - 2\text{CS}[\Re(\ve{s}), \Re(\tilde{\ve{s}})]}$.}.
We first point out that the power spectrum of $\ve{d}$ coincides with the sum of those of $\ve{s}$ and $\ve{n}$ because of the statistical independence between $\ve{s}$ and $\ve{n}$. The power spectra of $\Re(\ve{\tilde{s}})$ and $\Re(\ve{s})$ are in very good agreement with each other up to $0.18~\text{px}^{-1}$, at which scale the noise power is ten times that of the signal. At smaller scales, where the noise dominates the signal even more, our algorithm is not able to retrieve the true power spectrum but gets closer to it. The power spectrum of $\Re(\ve{d} - \ve{\tilde{s}})$ coincides with that of $\Re(\ve{n})$ at small scales, and this agreement progressively worsens toward larger scales. This large-scale behavior shows that our algorithm does not remove the already negligible noise, as shown by the superposition of the power spectrum of $\Re(\ve{s}-\ve{\tilde{s}})$ and that of $\Re(\ve{n})$ at these scales. The cross-spectrum $\tilde{\ve{s}}\times\ve{s}$ is slightly below the power spectrum of $\ve{s}$ at intermediate scales and this discrepancy increases toward the smallest scales.
At intermediate scales, where the power spectrum of $\Re(\ve{\tilde{s}})$ matches that of $\Re(\ve{s})$, we suspect this discrepancy to stem from differences between the phases of the Fourier components of $\ve{s}$ and $\tilde{\ve{s}}$ that would deserve a further quantification.
Nevertheless, the production of a denoised map whose power spectrum coincides with that of $\ve{s}$, even though $\ve{n}$ is ten times more powerful than $\ve{s}$, and that retains a significant correlation with $\ve{s}$ is a striking success of our method.

To better characterize the non-Gaussianity of $\tilde{\ve{s}}$, we computed the PDFs of the increments of $Q$ for the noisy, denoised, and truth maps, and we plot them in Fig.~\ref{fig:simulated_increments} for three scalar lags. The increment $\delta Q_l(\ve{r}$) for a scalar lag $l$ at a position $\ve{r}$ is the set of differences ${\delta Q_\ve{l}(\ve{r}) = Q(\ve{r}) - Q(\ve{r} + \ve{l})}$ with $l \leq |\ve{l}| < l + 1$ in pixel units.
These statistics are usually computed on velocity maps for the tails of the PDFs that characterize the intermittent dissipation of turbulence in the ISM~\citep{Hily-Blant2008, Hily-Blant2009}. Contrary to the case of the noisy map, the distributions of increments for $\Re(\ve{s})$ are far from Gaussian for every lag. This is a clear signature of the non-Gaussianity of the data as we expect Gaussian-distributed increments for homogeneous Gaussian data. Our method recovers these statistics and  their non-Gaussian tails with limited distortion for each lag, demonstrating its efficiency in retrieving non-Gaussianity in the data. A more thorough characterization of this non-Gaussianity has been postponed for further studies.

%%%%%%%%%%%%%%%%%%%%%%%%%%%%%%%%%%%%%%%%%%%%%%%%%%%%
%%%%%%%%%%%%%%%%%%%%%%%%%%%%%%%%%%%%%%%%%%%%%%%%%%%%
%% SECTION 3: APPLICATION TO PLANCK POLARIZATION DATA
%%%%%%%%%%%%%%%%%%%%%%%%%%%%%%%%%%%%%%%%%%%%%%%%%%%%
%%%%%%%%%%%%%%%%%%%%%%%%%%%%%%%%%%%%%%%%%%%%%%%%%%%%

\section{Application to \textit{Planck} polarization data}
\label{sec:planck}

\subsection{Presentation of the data}

We now apply our denoising method to a $Q+iU$ polarization map of the Chamaeleon-Musca region observed at 353~GHz with the {\it Planck} satellite~\citep[PR3 data\footnote{\url{https://wiki.cosmos.esa.int/planck-legacy-archive/index.php/Main_Page}},][]{planck2016-l03}. At this frequency, the CMB is negligible and the dominant components are the dust emission and the noise~\citep[see Fig.~34 of][]{planck2016-l04}. We consider the $Q$ and $U$ maps corresponding to the full mission, and those corresponding to the two half-missions, as well as the 300 end-to-end simulated $Q$ and $U$ maps of the noise and the systematics of the instrument for the full mission (the ones used in Sect.~\ref{sec:method}). We projected all of these maps on $512\times512$ grids with a pixel size of $2.35\arcmin$, centered on the region of the Chamaeleon-Musca clouds at Galactic coordinates ${(l, b) = (300.26^\circ, -16.77^\circ)}$. We used a Gnomonic projection through the {\tt HEALPix/healpy}\footnote{\url{http://healpix.sourceforge.net}} package~\citep{2005ApJ...622..759G, Zonca2019}.

In Fig.~\ref{fig:observed_maps}, we show the projected full-mission $Q$ map that we aim to denoise (left), as well as the denoised map discussed in the next section (middle). The same maps for $U$ are given in Fig.~\ref{fig:observed_maps_U} in Appendix~\ref{app:comparison}.

\subsection{Denoising results}

\begin{figure}
    \centering
    \includegraphics[width=0.95\hsize]{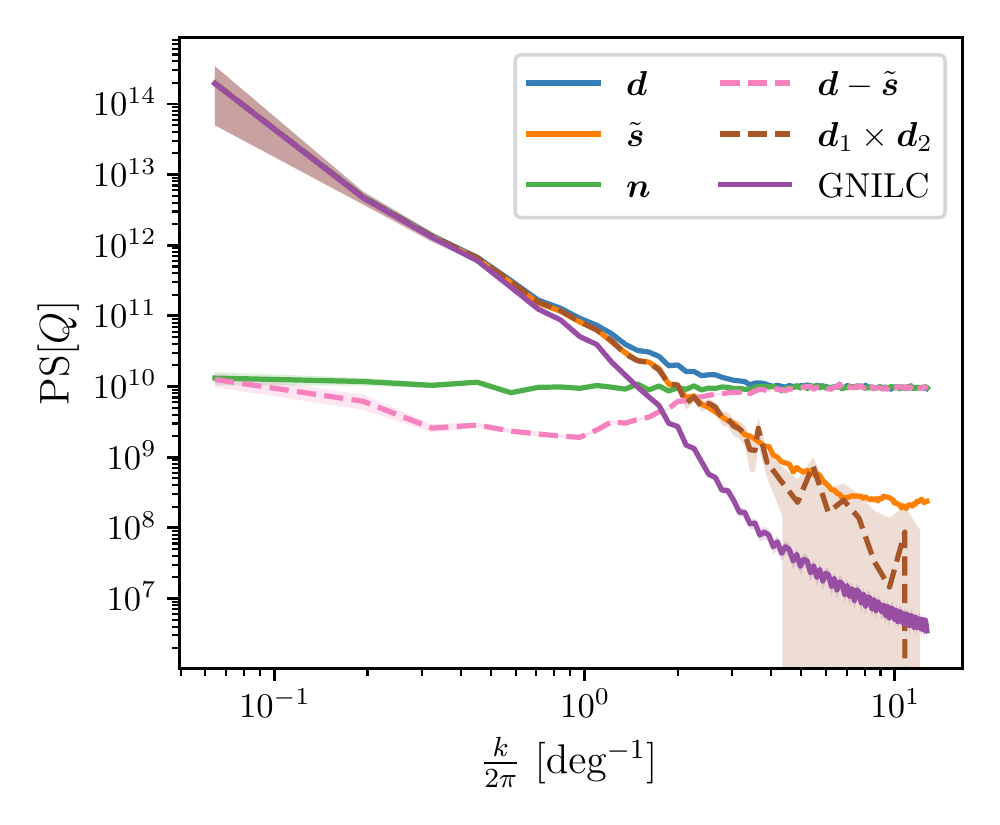}
    \caption{$Q$ maps power spectra for $\ve{d}$, $\ve{\tilde{s}}$, $\ve{n}$ (all in solid lines), and $\ve{d}-\ve{\tilde{s}}$, and cross-spectrum between $\ve{d_1}$ and $\ve{d_2}$ (both in dashed lines). This cross-spectrum estimates the power spectrum of the dust emission. Also shown in a solid line is the power spectrum of the GNILC $Q$ map, corresponding to the same Chamaeleon-Musca field.}
    \label{fig:observed_ps}
\end{figure}

\begin{figure*}
    \centering
    \includegraphics[width=\textwidth]{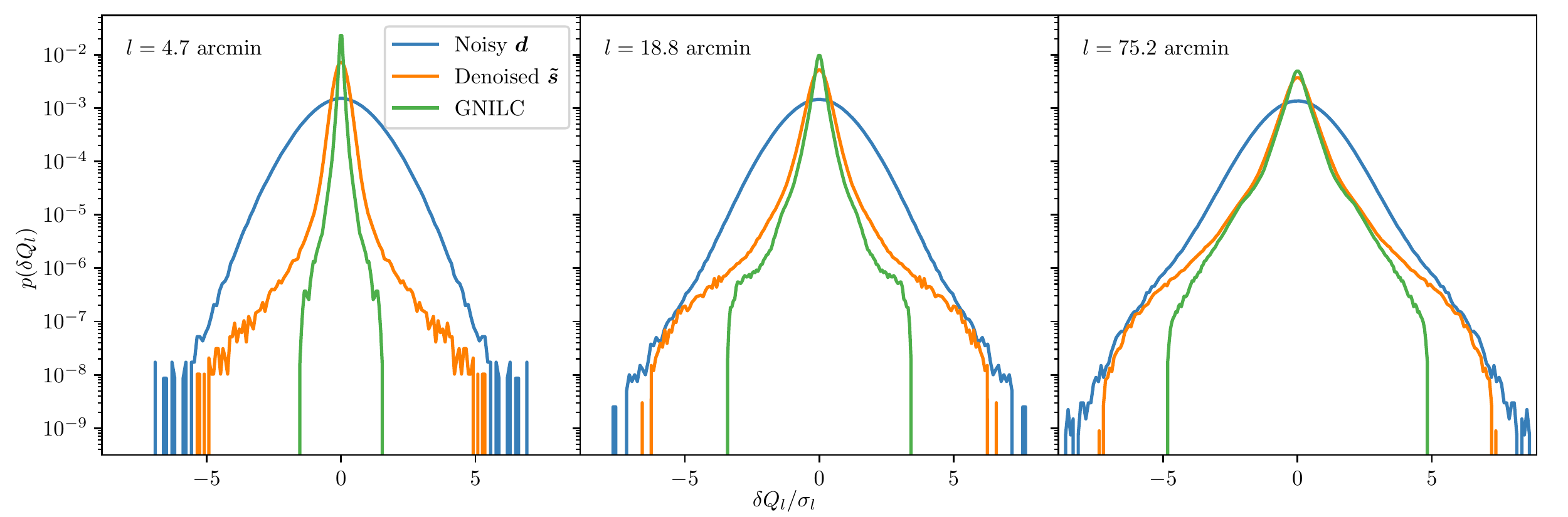}
    \caption{PDFs of the Chamaeleon-Musca $Q$ maps increments for $\ve{d}$ (noisy), $\ve{\tilde{s}}$ (denoised), and GNILC map, computed for three logarithmically spaced lags ranging from $4.7\arcmin$ to $75.2\arcmin$. For each lag, the increments were normalized by the standard deviation $\sigma_l$ of the Gaussian that fits the core of the PDF of the noisy map.}
    \label{fig:observed_increments}
\end{figure*}

We applied the denoising method presented in Sect.~\ref{sec:method} to the full mission $Q+iU$ map, using the corresponding 300 noises. Figures~\ref{fig:observed_maps} and \ref{fig:observed_maps_U} (middle) show the resulting denoised $Q$ and $U$ maps, respectively. The overall noise level has been clearly mitigated although subtle residuals of the patterns due to the scanning strategy remain, and we can now discern a more complex layout of structures even in the regions where the signal is weak.

In Fig.~\ref{fig:observed_ps}, we show a power spectrum analysis of the denoising of the $Q$ map, comparing the power spectra of the noisy and denoised maps $\Re(\ve{d})$ and $\Re(\tilde{\ve{s}})$, of one noise map $\Re(\ve{n})$, and of the difference between the noisy and denoised maps $\Re(\ve{d} - \tilde{\ve{s}})$, as well as the cross-spectrum between the two half-missions maps. This cross-spectrum gives an estimate for the power spectrum of the truth signal since the noise signals are independent. It is satisfactory to see that the power spectrum of $\Re(\tilde{\ve{s}})$ is consistent with this cross-spectrum for scales down to $k/(2\pi) \sim 3.4~\text{deg}^{-1}$. Also, $\Re(\ve{d} - \tilde{\ve{s}})$ behaves similarly to what we have observed in Sec.~\ref{sec:fig_merit}, with a power spectrum consistent with that of $\Re(\ve{n})$ when the noise has a larger power than the signal, and it falls below what is expected when the emission of the dust begins to dominate. Similar results are obtained for $U$.

Figure~\ref{fig:observed_increments} compares the PDFs of the increments of $Q$ for the noisy and denoised maps, and for the same three scalar lags as in Fig.~\ref{fig:simulated_increments}. Based on the results of Sect.~\ref{sec:method}, we expect the PDFs of the denoised map to be good estimates for the statistics of the truth map. These results show important differences between the noisy and denoised maps that are again a signature of the non-Gaussianity of the truth signal. We find similar results for the increments of $U$.

The literature on image denoising is rich and abundant~\citep[see for instance][]{Buades2010, Dabov2007, Zhang2017}, and a thorough comparison of our method with other denoising algorithms would be beyond the scope of this paper. Nevertheless, to give some elements of comparison, we compare our method in Appendix~\ref{app:comparison} to the following: (1)  Wiener filtering and sampling methods, which are widely used in astrophysics; and (2) the GNILC method~\citep{planck2016-XLVIII}, which was used on {\it Planck} polarization data and provides a local smoothing kernel in order to optimally remove the noise. For the Wiener approach, we find that neither the Wiener filtered image nor realizations drawn from the Wiener posterior distribution are able to retrieve the PDFs of the increments, even when the true power spectrum was given as input. The comparison with GNILC shows that our method recovers the true power spectrum and the PDFs of the increments better.

%%%%%%%%%%%%%%%%%%%%%%%%%%%%%%%%%%%%%%%%%%%%%%%%%%%%
%%%%%%%%%%%%%%%%%%%%%%%%%%%%%%%%%%%%%%%%%%%%%%%%%%%%
%% SECTION 4: CONCLUSION AND PROSPECTS
%%%%%%%%%%%%%%%%%%%%%%%%%%%%%%%%%%%%%%%%%%%%%%%%%%%%
%%%%%%%%%%%%%%%%%%%%%%%%%%%%%%%%%%%%%%%%%%%%%%%%%%%%

\section{Conclusion and prospects}
\label{sec:conclusion}

We have introduced a new method for the denoising of {\it Planck} polarization data based on the generation of random synthetic maps from WPH statistics that characterize the non-Gaussianity of the dust emission. This method takes advantage of the strong statistical differences between the signal of interest (non-Gaussian and regular) and the noise (close to Gaussian and irregular) by performing an optimization that constrains the statistical properties of the denoised map plus noise realizations.

We applied our method to mock $Q+iU$ noisy data designed to emulate typical {\it Planck} polarization maps of dust in the diffuse ISM. The denoised map has a power spectrum that coincides with the true power spectrum down to a minimum scale where the power of the noise is ten times that of the signal, while being highly correlated with the truth signal. It recovers the PDFs of the increments for various isotropic lags, demonstrating that our method is able to retrieve non-Gaussianity in the data. Finally, we applied this method to a 353~GHz {\it Planck} observation of the Chamaeleon-Musca field.

Our method has been introduced as a statistical denoising method, but it should be more generally applicable to component separation problems. In particular, we expect this method to be efficient at disentangling  dust emission from the CMB, cosmic infrared background, and noise in {\it Planck} maps, provided that we have relevant models of each of these contaminants at our disposal. Therefore, it could hopefully enhance the scientific outcome of {\it Planck} data.

One of the main motivations of our work is to build a generative model of the dust polarization signal that takes into account the non-Gaussianity of the data. Such a model derived from the analysis of {\it Planck} data may be used to simulate the dust polarization sky. Our denoising algorithm constitutes a step toward this modeling, in the sense that WPH statistics of the dust emission, corrected to a first approximation from noise contamination, may be computed from the denoised map. A natural extension of this work would be to quantitatively assess how well we can recover the WPH statistics of the truth map with this approach. This will be the subject of a future paper.

\begin{acknowledgements}
We thank the anonymous referee for a very helpful report.
This work was undertaken in parallel to a similar project led by J.-M. Delouis on the full sky, and we acknowledge a fruitful interaction with his team. We gratefully acknowledge P. Lesaffre for helping us with the computation of the PDFs of the increments. We also acknowledge fruitful discussions with J.-F. Cardoso, S. Mallat, and T. Marchand. This research was supported by the Agence Nationale de la Recherche (project BxB: ANR-17-CE31-0022). B. Regaldo-Saint Blancard acknowledges support from the Centre National d'Etudes Spatiales (CNES).
\end{acknowledgements}

% For the bibliography, at the end
\bibliographystyle{bibtex/aa} % style aa.bst
\bibliography{bibtex/bib.bib,bibtex/Planck_bib.bib}

\begin{appendix}

%%%%%%%%%%%%%%%%%%%%%%%%%%%%%%%%%%%%%%%%%%%%%%%%%%%%
%%%%%%%%%%%%%%%%%%%%%%%%%%%%%%%%%%%%%%%%%%%%%%%%%%%%
%% APPENDIX 1: ?
%%%%%%%%%%%%%%%%%%%%%%%%%%%%%%%%%%%%%%%%%%%%%%%%%%%%
%%%%%%%%%%%%%%%%%%%%%%%%%%%%%%%%%%%%%%%%%%%%%%%%%%%%

\section{The WPH statistics}
\label{app:wph}

In this appendix, we introduce the WPH statistics used in this paper, referring to \citet{Allys2020} and \citet{Mallat2020} for additional details. We also describe the detailed denoising procedure introduced in Sect.~\ref{sec:desc_method}. We call $\ve{X}$ a given random field and $\ve{x}$ one of its realizations.

\subsection{Bump-steerable wavelets}

The WPH statistics rely on a bank of wavelets that are dilations and rotations of a mother bump-steerable wavelet $\ve{\psi}$ defined in Fourier space as follows~\citep{Mallat2020}:
\begin{multline}
    \hat{\ve{\psi}} (\vec k) = \exp \left(\frac{-(k - \xi_0)^2}{\xi_0^2 - (k-\xi_0)^2}\right) \cdot \mathbf{1}_{[0, 2\xi_0]}(k)  \\
    \times \cos^{L-1} (\arg (\vec k)) \cdot \mathbf{1}_{[0, \pi/2]}(|\arg (\vec k)|),
\end{multline}
with $k = \norm{\ve{k}}$, $\mathbf{1}_A(x)$ being the indicator function that returns 1 if $x\in A$ and $0$ otherwise, and $\xi_0 = 0.85\pi$ being the central wavenumber of the mother wavelet. The dilated and rotated wavelets are defined as follows:
\begin{equation}
    \ve{\psi}_{\ve{\xi}_{j, \theta}}(\ve{r}) = 2^{-j}\ve{\psi}\left(2^{-j}\text{rot}_{-\theta}[\ve{r}]\right),
\end{equation}
with $j$ being the index of the dilation by a factor $2^j$ and $\theta$ being the angle of rotation. We call $\ve{\xi}_{j, \theta} = 2^{-j}\xi_0\ve{u}_\theta$ with ${\ve{u}_\theta = \cos\theta~\ve{u}_x+\sin\theta~\ve{u}_y}$ the central wavevector of the wavelet obtained by such a transform. In practice, we consider $J$ dilation indices $j$ ranging from 0 to $J-1$, and we divide $2\pi$ into $2L$ evenly spaced rotation angles $\theta$. The bank of wavelets is thus the set of $2JL$ wavelets ${\{\ve{\psi}_{\ve{\xi}_{j,\theta}} : j=0, \dots, J - 1, \theta=0, \frac{\pi}{L}, \dots,\frac{(2L-1)\pi}{L}\}}$, and these wavelets cover most of Fourier space with their respective band-passes.

\begin{figure}
    \centering
    \includegraphics[width=\hsize]{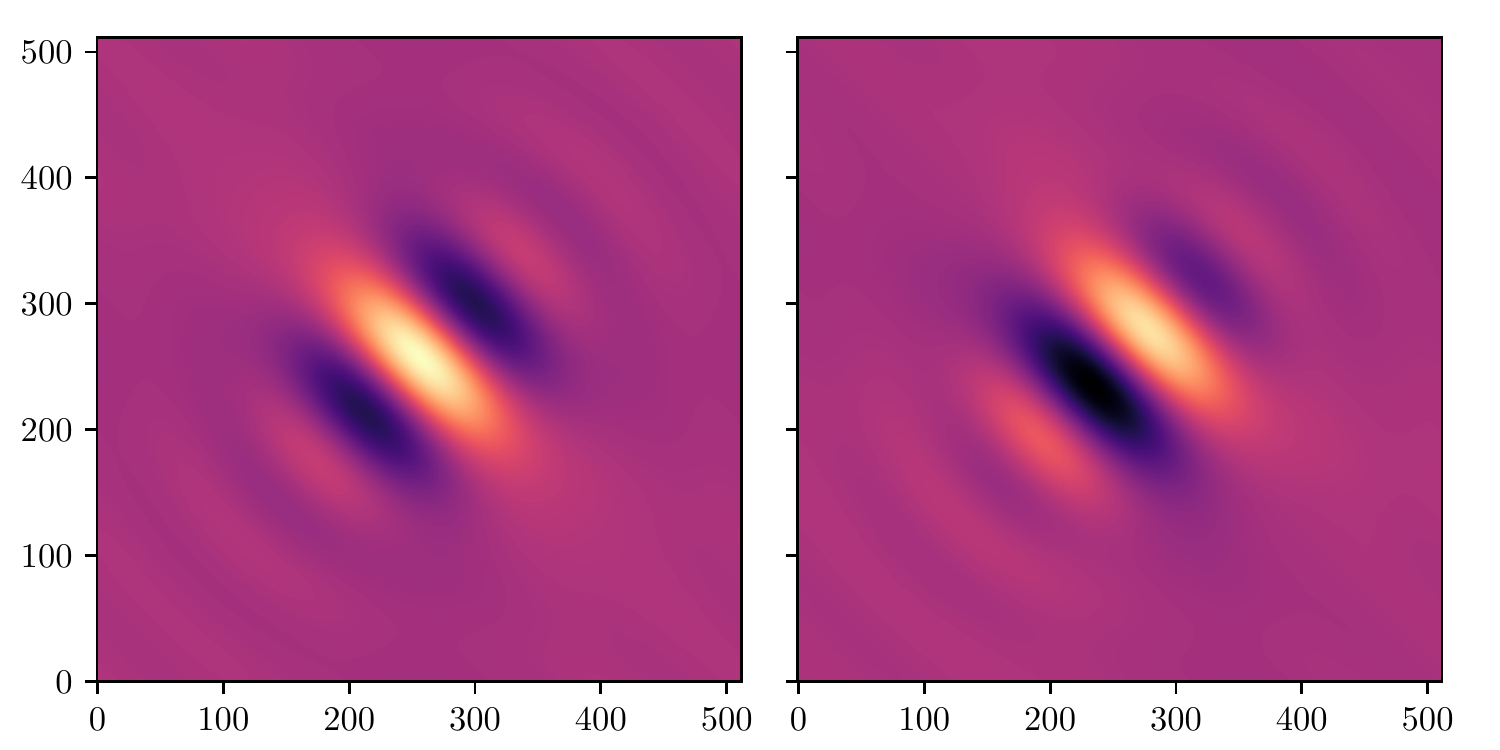}
    \caption{Real part (left) and imaginary part (right) of a $512\times 512$ bump-steerable wavelet $\ve{\psi}_{6, \pi/4}$. The wavelet is centered in the middle of the map for a better visualization.}
    \label{fig:wavelet}
\end{figure}

In this paper, we work with $512\times 512$ maps and choose ${J = 8}$ and ${L = 8}$. We show in Fig.~\ref{fig:wavelet} one example wavelet from the bank.

\subsection{WPH moments}

The WPH moments of $\ve{X}$ are covariances of the phase harmonics of the wavelet transform of $\ve{X}$. With $\{\ve{\psi}_{\ve{\xi}_1}, \ldots, \ve{\psi}_{\ve{\xi}_N}\}$, our bank of wavelets labeled by their central wavevectors $\ve{\xi}_i$, the wavelet transform of $\ve{X}$ corresponds to the set $\{\ve{X} \star \ve{\psi}_{\ve{\xi}_1}, \dots, \ve{X}\star \ve{\psi}_{\ve{\xi}_N}\}$, where $\star$ denotes the convolution operation. These convolutions amount to a local band-pass filtering of $\ve{X}$ on the scales probed by each of the wavelets. For a homogeneous $\ve{X}$, the WPH moments of $\ve{X}$ are as follows:
\begin{equation}
C_{\ve{\xi}_i,p_i,\ve{\xi}_j,p_j}(\ve{\tau}) = \text{Cov} \left(\left[\ve{X} \star \ve{\psi}_{\ve{\xi}_i}(\ve{r})\right]^{p_i},\left[\ve{X} \star \ve{\psi}_{\ve{\xi}_j}(\ve{r} + \ve{\tau})\right]^{p_j}\right),
\end{equation}
with $z \mapsto \left[z \right]^p = |z| \cdot \text{e}^{i p~\text{arg}(z)}$ being the phase harmonic operator\footnote{These moments do not depend on the $\ve{r}$ variable because of the homogeneity of $\ve{X}$.}.

These WPH moments are able to capture interactions between different scales of $\ve{X}$ thanks to the phase harmonic operator. Indeed, the covariance between $\ve{X} \star \ve{\psi}_{\ve{\xi}_i}$ and $\ve{X} \star \ve{\psi}_{\ve{\xi}_j}$ vanishes when the wavelets $\ve{\psi}_{\ve{\xi}_i}$ and $\ve{\psi}_{\ve{\xi}_j}$ have nonintersecting band-passes, and it is otherwise a function of the power spectrum of $\ve{X}$ only. With proper $p_i$ and $p_j$ indices, the phase harmonic operator can make $[\ve{X} \star \ve{\psi}_{\ve{\xi}_i}]^{p_i}$ and $[\ve{X} \star \ve{\psi}_{\ve{\xi}_j}]^{p_j}$ comparable in the sense that they share common spatial frequencies, allowing an extraction of high-order information through their covariance.

We discretize the $\tau$ variable similarly to \citet{Allys2020}. This introduces a grid of $\ve{\tau}_{n, \alpha}$ vectors for each wavelet $\ve{\psi}_{j, \theta}$, defined as
\begin{equation}
    \ve{\tau}_{n, \alpha} = 3n2^j\ve{u}_{\theta + \alpha},
\end{equation}
with $n = 0, \dots, \Delta_n$ and $\alpha\in \{-\pi/4, 0, \pi/4, \pi/2\}$. To avoid a redundancy in the information contained in the coefficients, we discarded the translations for which $n > \min(J-1-j, \Delta_n)$.

\citet{Allys2020} identified a relevant set of WPH moments to build a generative model of real-valued simulated data of the large-scale structure of the Universe, which are highly non-Gaussian at late times and at scales smaller than 100 $h^{-1}$Mpc. In the present work, we have adopted the same corresponding WPH statistics and applied them similarly on complex-valued data, except for the scaling moments discussed below. In practice we estimate a set of moments that can be divided into the following five categories:
\begin{itemize}
    \item the $S^{(1,1)}$ moments, of the form $C_{\ve{\xi},1,\ve{\xi},1}$, at every $\ve{\tau}_{n, \alpha}$ with $\Delta_n = 2$. They capture the power spectrum information of the $\ve{\xi}$ band-pass.
    \item the $S^{(0,0)}$ moments, of the form $C_{\ve{\xi},0,\ve{\xi},0}$, at every $\ve{\tau}_{n, \alpha}$ with $\Delta_n = 2$. They capture information related to the sparsity of the data in the $\ve{\xi}$ band-pass.
    \item the $S^{(0,1)}$ moments, of the form $C_{\ve{\xi},0,\ve{\xi},1}$, at $\ve{\tau} = 0$ only. They capture information related to the couplings between the scales of the same $\ve{\xi}$ band-pass.
    \item the $C^{(0,1)}$ moments, of the form $C_{\ve{\xi}_1,0,\ve{\xi}_2,1}$, considering ${0 \leq j_1 < j_2 \leq J - 1}$ and ${\lvert\theta_1-\theta_2\rvert\leq \frac{\pi}{2}}$, at every $\ve{\tau}_{n, \alpha}$ with ${\Delta_n = 2}$ when $\theta_1 = \theta_2$ and at $\ve{\tau} = 0$ when $\theta_1 \neq \theta_2$. They capture information related to the correlation between local levels of oscillation for the scales in the $\ve{\xi}_1$ and $\ve{\xi}_2$ band-passes.
    \item the $C^{\rm phase}$ moments, of the form $C_{\ve{\xi}_1,1,\ve{\xi}_2, p_2}$ with ${p_2 = \xi_1/\xi_2 > 1}$, considering $0 \leq j_1 < j_2 \leq J - 1$ and $\theta_1 = \theta_2$, at every $\ve{\tau}_{n, \alpha}$ with $\Delta_n = 2$. They capture information related to the statistical phase alignment of oscillations between the scales in the $\ve{\xi}_1$ and $\ve{\xi}_2$ band-passes.
\end{itemize}

We call $\tilde{C}_{\ve{\xi}_1,p_1,\ve{\xi}_2,p_2}(\ve{\tau})$ the normalized estimates of the WPH moments $C_{\ve{\xi}_1,p_1,\ve{\xi}_2,p_2}(\ve{\tau})$. In practice they are defined as follows for a map $\ve{x}$:
\begin{equation}
    \tilde{C}_{\ve{\xi}_1,p_1,\ve{\xi}_2,p_2}(\ve{\tau}) = \frac{\big \langle\ve{x}^{(\ve{\xi}_1, p_1)}\left(\ve{r}\right)\overline{\ve{x}^{(\ve{\xi}_2, p_2)}}\left(\ve{r}+\ve{\tau}\right)\big \rangle}{\sqrt{\big \langle\lvert{\ve{x}_0^{(\ve{\xi}_1, p_1)}}\rvert^2\big \rangle\big \langle\lvert{\ve{x}_0^{(\ve{\xi}_2, p_2)}}\rvert^2\big \rangle}},
\end{equation}
where the brackets stand for a spatial mean on the $\ve{r}$ variable, the bar denotes the complex conjugate, and ${\ve{x}^{(\ve{\xi}, p)} = [\ve{x} \star \ve{\psi}_{\ve{\xi}}]^{p} - \big \langle[\ve{x}_0 \star \ve{\psi}_{\ve{\xi}})]^{p}\big \rangle}$. These estimates depend on a reference map $\ve{x}_0$. We found that the choice of $\ve{x}_0$ can have a significant impact on the results of our denoising method as explained below.

\subsection{Scaling moments}

Similarly to \citet{Allys2020}, we complement these statistics with a small number of coefficients, the estimates of the so-called scaling moments $L_{j, p}$ that better constrain the largest scales that are not probed by the WPH moments. These moments were constructed from a bank of scaling functions $\{\varphi_j\}_{0 \leq j \leq J-1}$, which correspond to dilations of an isotropic Gaussian filter $\varphi$ defined in Fourier space by the following:
\begin{equation}
 \hat \varphi (\vec k) = \exp \left(-\frac{||\vec k||^2}{2 \sigma^2} \right),
\end{equation}
with $\sigma=0.496\times 2^{-0.55} \xi_0$.
For a real-valued random field $\ve{X}$, the scaling moments of $\ve{X}$ are as follows:
\begin{align}
L_{j,0} &= \text{Cov} \left[\lvert\ve{X} \star \varphi_{j} \rvert,\lvert\ve{X} \star \varphi_j \rvert \right],\\
L_{j, p} &= \text{Cov} \left[\left(\ve{X} \star \varphi_{j} \right)^{p},\left(\ve{X} \star \varphi_j \right)^{p} \right]~(\text{for}~p>0).
\end{align}
In practice, we only estimated the scaling moments with ${p\in\{0, 1, 2, 3\}}$ and $2\leq j \leq J - 2$. We define the corresponding normalized estimates computed for a map $\ve{x}$ by
\begin{align}
    \tilde{L}_{j,p} = \frac{\langle \lvert \ve{x}^{(j,p)}\rvert^2 \rangle}{\langle \lvert \ve{x}_0^{(j,p)}\rvert^2 \rangle},
\end{align}
with
\begin{align}
    \ve{x}^{(j,0)} &= \lvert \left(\ve{x} - \langle\ve{x}_0\rangle\right)\star \varphi_j \rvert - \langle  \lvert \left(\ve{x} - \langle\ve{x}_0\rangle\right)\star \varphi_j \rvert \rangle, \\
     \ve{x}^{(j,p)} &= \left( \left(\ve{x} - \langle\ve{x}_0\rangle\right)\star \varphi_j \right)^p - \langle  \left( \left(\ve{x} - \langle\ve{x}_0\rangle\right)\star \varphi_j \right)^p \rangle ~(\text{for}~p>0).
\end{align}
These normalized estimates depend on the same reference map $\ve{x}_0$ introduced for the estimates of the WPH moments of $\ve{X}$.
As we deal with complex maps $\ve{x}$ in this work, we computed these estimates for their real and imaginary parts separately.

\subsection{Denoising procedure}

We applied our denoising method to a simulated noisy map $\ve{d}$ in two steps. We performed a first denoising of $\ve{d}$ using a $\phi$ operator that only takes into account the estimates of the $S^{(1,1)}$ moments, related to the power spectrum information, and the estimates of the scaling moments, and by normalizing the estimates choosing $\ve{u}_0 = \ve{d}$. This first step yields a map $\ve{\tilde{s}}_0$ that has a power spectrum that is much more consistent with the truth map $\ve{s}$.
Then, we performed a second denoising of $\ve{d}$ using a $\phi$ operator that includes the whole set of estimates of WPH moments and scaling moments, but here normalizing these estimates with $\ve{u}_0 = \ve{\tilde{s}}_0$. We call $\tilde{\ve{s}}$ the output of this second step. This particular choice for the normalization of the estimates has proven to be more efficient at retrieving the power spectrum and the PDFs of the increments of the truth map $\ve{s}$.

In Table~\ref{table:nb_coeffs}, we show the resulting number of statistical coefficients per class of moments computed for a complex $512\times512$ pixels map $\ve{x}$ and for the chosen parameters. In this paper, the $\phi$ operator used for the second step of the procedure computes a set of 11176 statistical coefficients (of which 40 correspond to the estimates of the scaling moments), which corresponds to approximately 4\% of the number of pixels of the input map.

\begin{table}
    \def\arraystretch{1.5}
    \centering
    \begin{tabular}{cccccc|c}
        \hline
        \hline
        $S^{(1,1)}$ & $S^{(0,0)}$ & $S^{(0,1)}$ & $C^{(0,1)}$ & $C^{\rm phase}$ & $L_{j,p}$ & Total \\
        \hline
           960 & 960 & 128 & 6336 & 2752 & 40 & 11176 \\ 
        \hline
    \end{tabular}
    \caption{Number of statistical coefficients per class of moments for the parameters of this paper.}
    \label{table:nb_coeffs}
\end{table}

%%%%%%%%%%%%%%%%%%%%%%%%%%%%%%%%%%%%%%%%%%%%%%%%%%%%
%%%%%%%%%%%%%%%%%%%%%%%%%%%%%%%%%%%%%%%%%%%%%%%%%%%%
%% APPENDIX 2: ?
%%%%%%%%%%%%%%%%%%%%%%%%%%%%%%%%%%%%%%%%%%%%%%%%%%%%
%%%%%%%%%%%%%%%%%%%%%%%%%%%%%%%%%%%%%%%%%%%%%%%%%%%%

\section{Comparison to other methods}
\label{app:comparison}

\begin{figure*}
    \centering
    \includegraphics[width=\textwidth]{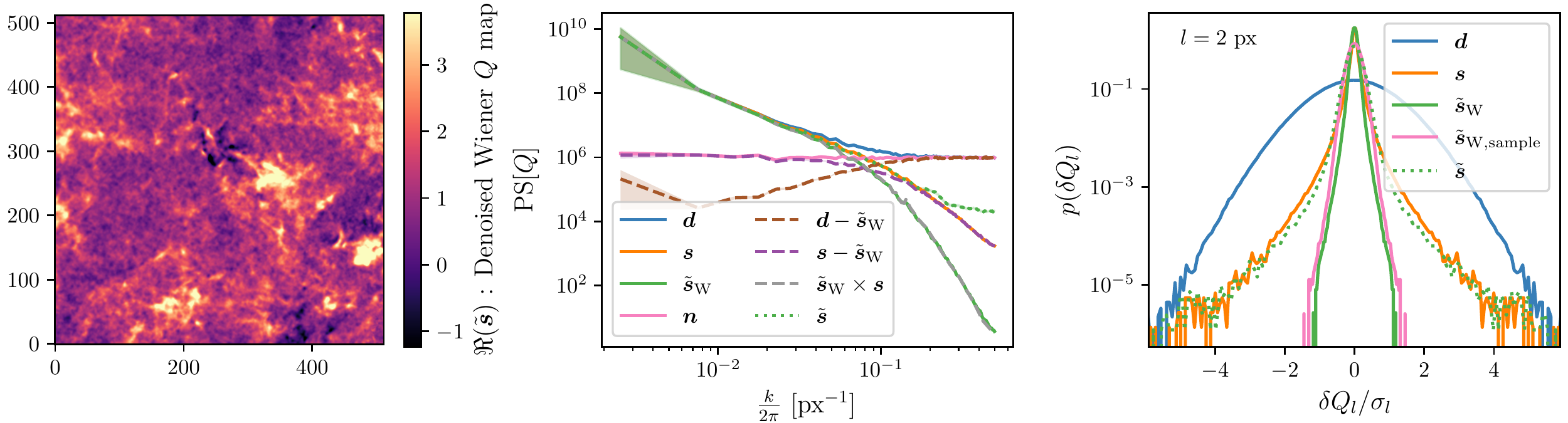}
    \caption{\textit{Left:} Wiener filtered map $\ve{\tilde{s}}_{\rm W}$. \textit{Middle:} Same as Fig.~\ref{fig:simulated_ps}, but for $\ve{\tilde{s}}_{\rm W}$. The power spectrum of $\ve{\tilde{s}}_{\rm W}$ is suppressed as expected. \textit{Right:} Same as Fig.~\ref{fig:simulated_increments}, but for $l = 2 \mathrm{px}$ with the Wiener filter (mean) map and a Wiener posterior sample.}
    \label{fig:simulated_wiener}
\end{figure*}

\begin{figure*}
    \centering
    \includegraphics[width=0.95\textwidth]{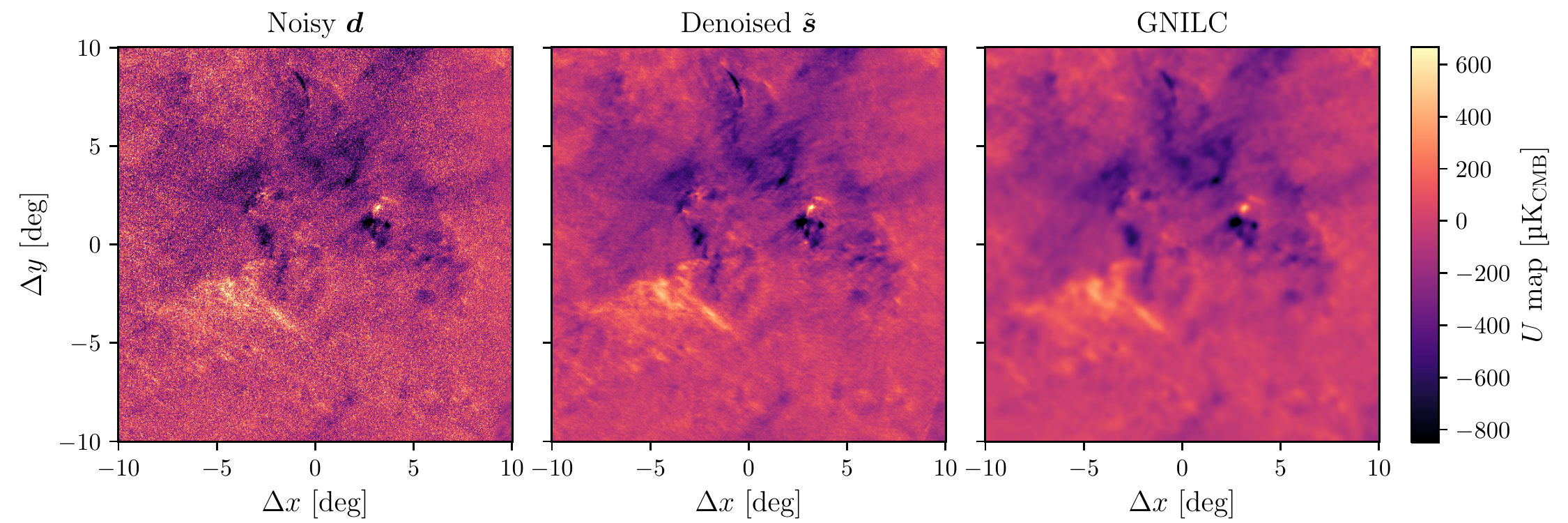}
    \caption{Same as Fig.~\ref{fig:observed_maps}, but for $U$.}
    \label{fig:observed_maps_U}
\end{figure*}

\subsection{Wiener filtering and posterior sampling}

Wiener filtering is a typical approach to signal inference for problems of the form given by Eq.~\ref{eq:data_model}. In this case, the Wiener filtered reconstruction $\tilde{\mathbf{s}}_W$ (\citealt{wiener1949extrapolation},~\citealt{zaroubi_wiener}) is given by
\begin{equation} \label{eq:wiener_filter}
\tilde{\mathbf{s}}_W = \mathbf{W} \mathbf{d} = \mathbf{S}  \big( \mathbf{S}  + \mathbf{N} \big)^{-1} \mathbf{d}  \ ,
\end{equation}
where $\mathbf{S} = \langle \mathbf{s} \mathbf{s}^\dagger \rangle$ and $\mathbf{N} = \langle \mathbf{n} \mathbf{n}^\dagger \rangle$ are the signal and noise covariance matrices, respectively. For brevity we have assumed $\langle \mathbf{s} \rangle = \langle \mathbf{n} \rangle = 0$. This is the linear filter that gives the expected minimum variance of residuals $(\mathbf{Wd} - \mathbf{s})$ for known $\mathbf{S}$ and $\mathbf{N}$. 

Furthermore, it is both the maximum a posteriori and mean posterior solution if the noise and signal are drawn from Gaussian distributions. That is, if the real-valued data are distributed with a likelihood 
\begin{equation}
p( \mathbf{d} | \mathbf{s}, \mathbf{N} ) = \frac{1}{\sqrt[]{(\mathrm{det} 2 \pi \mathbf{N})}} \mathrm{exp} \Big[ - \frac{1}{2} ( \mathbf{d} -  \mathbf{s} )^\dagger \mathbf{N}^{-1} ( \mathbf{d} -  \mathbf{s} ) \Big] \ ,
\end{equation}
and the prior distribution for the real-valued signal is
\begin{equation}
p( \mathbf{s} | \mathbf{S} ) = \frac{1}{\sqrt[]{(\mathrm{det} 2 \pi \mathbf{S})}} \mathrm{exp} \Big[ - \frac{1}{2} \mathbf{s}^\dagger \mathbf{S}^{-1} \mathbf{s}  \Big] \ ,
\end{equation}
then the Wiener posterior distribution is given by
\begin{equation} \label{eq:wienerpost}
\begin{split}
p( \mathbf{s} | \mathbf{S}, \mathbf{N}, \mathbf{d} )  &\propto  p( \mathbf{d} | \mathbf{s}, \mathbf{N} ) p( \mathbf{s} | \mathbf{S} )  \\
 &\propto \mathrm{exp} \Big[  - \frac{1}{2} ( \mathbf{s} - \mathbf{W} \mathbf{d} )^\dagger (  \mathbf{S}^{-1} +  \mathbf{N}^{-1}  ) (\mathbf{s} - \mathbf{W} \mathbf{d}) \Big] \ .
\end{split}
\end{equation}
\noindent For the problem described in this work, the noise covariance $\mathbf{N}$ can be estimated from the samples of noise realizations and the signal covariance $\mathbf{S}$ can either be assumed or jointly estimated from the data \citep{wandelt_sampling}.

We avoided inverting the dense $\mathbf{W}$ matrix by implementing a messenger field approach, first described by~\cite{elsner2013} and now widely used in cosmology (e.g., \citealt{jasche15, alsing2017cosmological, jeffrey2018, wiener_dfe, 10.1093/mnras/stz2608}). We assumed that the underlying signal is homogeneous and isotropic, such that the signal covariance $\mathbf{S}$ is diagonal in harmonic space with elements corresponding to the one-dimensional power spectrum. The uncorrelated noise gives a diagonal noise covariance $\mathbf{N}$ in pixel space, so that the messenger field can efficiently iterate between harmonic and pixel space.

As we are concerned with polarization data, we used a \mbox{spin-2} harmonic transformation between $\{\mathbf{s}_Q, \mathbf{s}_U \}$ and $\{\widetilde{\mathbf{s}}_E, \widetilde{\mathbf{s}}_B \}$, using the curl-free E-mode and divergence-free B-mode representation. The signal covariance in harmonic space is therefore a concatenation of $\{{\mathbf{S}}_E, {\mathbf{S}}_B \}$. This formulation preserves the relevant $Q$-$U$ correlation.

Even if the Wiener filtered reconstructed signal $\tilde{\mathbf{s}}_W$ is the maximum of the posterior distribution $p( \mathbf{s} | \mathbf{d} )$, functions $f(\tilde{\mathbf{s}}_W)$ do not correspond to the maximum $p( f(\mathbf{s}) | \mathbf{d} )$. For example, the two-point statistics (e.g., variance or power spectra) of $\tilde{\mathbf{s}}_W$ are not unbiased estimates of the power spectrum of ${\mathbf{s}}$. Instead, we can draw sample images ${\mathbf{s}}_i$  from the posterior $p( \mathbf{s} | \mathbf{d} )$ (Eq.~\ref{eq:wienerpost}), so that the transformed samples $f({\mathbf{s}}_i)$ are correctly distributed according to $p( f(\mathbf{s}) | \mathbf{d} )$. These realizations are generated by amending the messenger field algorithm~\citep{elsner2013}.

Fig.~\ref{fig:simulated_wiener} shows the Wiener filtered reconstruction of the simulated $Q$ map (\textit{left panel}). The low amplitude power spectrum of the residual map $\Re(\ve{s} - \tilde{\ve{s}}_W)$ demonstrates that the pixel values are close to the truth, whereas the power spectrum of $\Re(\tilde{\ve{s}}_W)$ is biased low (\textit{center panel}), which is as expected. To compare the methods, we compared with functions of samples $\ve{s}_i$ from the Wiener posterior. Though not plotted here, as we input the true power spectrum, which was not done for the WPH denoising method, the realizations $\ve{s}_i$ did have the correct power spectra with relatively small sample variance.

As a goal of this denoising work is to retain the statistical non-Gaussianity in the signal, which could be represented in the WPH coefficients, we again compared the PDF of increments $\delta Q_l$. The right panel of~\ref{fig:simulated_wiener} shows the comparison between the data $\Re(\ve{d})$, the signal $\Re(\ve{s})$, the Wiener filtered (mean) map $\Re(\tilde{\mathbf{s}}_W)$, and a realization $\Re(\tilde{\ve{s}}_{W, \mathrm{sample}})$ drawn from the Wiener posterior. We see that neither the Wiener mean map $\Re(\tilde{\ve{s}}_W)$ nor the sample $\Re(\tilde{\ve{s}}_{W, \mathrm{sample}})$ capture the non-Gaussianity in terms of $\delta Q_l$ increments as successfully as our WPH denoising method. 
Indeed, for $\lvert \delta Q_l\rvert / \sigma_l \gtrapprox 1$, there is a clear divergence from the true PDF, showing that the tails of the true statistics are not recovered at all.

Furthermore, inspection of the WPH coefficients of the Wiener posterior samples $\phi(\ve{s}_i)$ shows them to be noticeably further from those of the truth map than $\phi(\tilde{\ve{s}})$. These preliminary results are encouraging, but not surprising. The Gaussian prior distribution in the statistical model of the Wiener posterior leads to a poor recovery of the non-Gaussian structures that are intrinsic to the polarized dust emission.

\subsection{GNILC}

The GNILC method is a wavelet-based component separation method that is designed to extract the emission of the Galactic foregrounds from the {\it Planck} full-sky maps. It has been applied to polarization maps $Q$ and $U$ \citep{planck2016-l04, planck2016-l11B} to disentangle the thermal dust polarization emission from the CMB polarization and instrumental noise over the entire sky. We show in Figs.~\ref{fig:observed_maps} and~\ref{fig:observed_maps_U} (right column) the resulting $Q$ and $U$ maps, respectively, for the Chamaeleon-Musca field observed at 353~GHz. We note that at this frequency, the CMB can be safely ignored, so that the main effect of the GNILC algorithm is to denoise the dust emission. In these maps, we clearly see that the smallest scales, most contaminated by the noise, have been filtered out. Compared to GNILC, our denoised maps thus include a wider variety of structures at intermediate and small scales.

A quantitative comparison based on the power spectrum and the PDFs of the increments for the $Q$ maps is shown in Figs.~\ref{fig:observed_ps} and \ref{fig:observed_increments}. The power spectrum of the GNILC map plummets at small scales and exhibits a lack of power compared to that of $\tilde{\ve{s}}$ and the cross-spectrum $\ve{d}_1\times\ve{d}_2$. The PDFs of the increments also show important discrepancies, especially at the tails of the PDFs, between the $\tilde{\ve{s}}$ map and the GNILC map. Assuming that the statistics of $\tilde{\ve{s}}$ give a relevant order of magnitude of the true statistics, based on the validation on simulated data, this suggests a distortion of these statistics by the GNILC method. This quantitative comparison demonstrates the superiority of our method for recovering the true power spectrum and, a priori, the PDFs of the increments.

\end{appendix}

\end{document}